\documentstyle[epsfig]{lamuphys}
\makeatletter
\let\chapter\hid@chapter
\makeatother
\begin{document}

\def\ltsima{$\; \buildrel < \over \sim \;$}
\def\simlt{\lower.5ex\hbox{\ltsima}}
\def\gtsima{$\; \buildrel > \over \sim \;$}
\def\simgt{\lower.5ex\hbox{\gtsima}}
\def\lya{Ly$\alpha$}

\newcommand{\minpoint}{\mbox{$'\mskip-4.7mu.\mskip0.8mu$}}

\pagenumbering{arabic}
\title{Element Abundances at High Redshifts{\normalsize $^1$}}
\author{Max\,Pettini\inst{}}

\institute{Institute of Astronomy, Madingley Road, Cambridge,
CB3 0HA, England}

\maketitle

\addtocounter{footnote}{+1}
{\footnotetext{{\tt Proceedings of ESO Workshop:} 
{\it Chemical Evolution from Zero to
High Redshift.} {\tt To be published in} {\it Lecture Notes in Physics},
{\tt ed. J Walsh \& M. Rosa, (Berlin: Springer).}}

\begin{abstract}
I review measurements of element abundances
in different components of the high redshift universe, including the 
Lyman alpha forest, damped Lyman alpha systems, and Lyman break galaxies.
Although progress is being made in all three areas, recent work
has also produced some surprises and shown that established ideas about 
the nature of the damped \lya\ systems in particular may be too simplistic.
Overall, our knowledge of metal abundances at high $z$ is still 
very sketchy. Most significantly, 
there seems to be an order of magnitude
shortfall in the comoving density of metals which have been measured up 
to now compared with those produced 
by the star formation activity seen in Lyman break 
galaxies. At least some of the missing metals are likely to be in 
hot gas in galactic halos and proto-clusters.
\end{abstract}

\section{Introduction}
The study of element abundances at large look-back times has grown 
considerably during the 1990s to encompass several broad themes  
including:

\begin{description}
	\item[a.]  {\bf Primordial Abundances of Light Elements.}~ One of the pillars 
	of the standard model, big-bang nucleosynthesis relates the 
	abundances of He, D and Li (relative to H) to $\Omega_B$,
	the density of baryons in the universe expressed as a fraction of the 
	closure density. The currently favoured value is 
	from the work by Burles \& Tytler (1998),
	$\Omega_Bh_{50}^2 = 0.076  \pm 0.004$
	($h_{50}$ is the Hubble constant in 
	units of 50~km~s$^{-1}$~Mpc$^{-1}$, assumed throughout),  
	but it is difficult to 
	know whether this error estimate can be fully trusted, given the paucity of 
	measurements of D/H and the systematic uncertainties which affect the 
	determination of the He abundance.
	Possibly the final word on $\Omega_B$ 
	will have to await the results of an entirely independent test, such 
	as the height of the first acoustic peak in the power 
	spectrum of the cosmic microwave background anisotropy (e.g. Turner 1999).
	
	\item[b.]  {\bf Census of Metals at Different Cosmic Epochs.}~
	Determining how the metal content of the universe increased with time
	is one aspect of the current concerted efforts to trace the global history 
	of star formation over the Hubble time. It is closely linked to issues 
	such as the measurement of the extragalactic background at different 
	wavelengths and the importance of dust in altering our view of 
	the high redshift 
	universe.

	\item[c.]  {\bf Element Ratios as a Function of Metallicity.}~
	By providing data which refer to different epochs 
	and different environments, 
	measurements at high redshift are a much needed---and yet to be  
	exploited---complement to local studies of the relative abundances of 
	different chemical elements in Galactic stars and H~II regions of nearby 
	galaxies. The ultimate purpose is the same as that which has motivated
	this field for the last thirty years: to test observationally our ideas 
	of the nucleosynthetic origin of different elements and search in the 
	pattern of relative abundances for clues to the nature of the first 
	episodes of metal enrichment.

	\item[d.]   {\bf Abundances in Active Galactic Nuclei.}~ 
	Here too chemical abundances, 
	if they can be determined reliably in such extreme 
	physical environments, hold  clues to the nature and timescales of the 
	AGN phenomenon.
	
\end{description}
At this meeting we saw each of these themes addressed by a variety of 
presentations. Given the limitations of space, 
in this review I shall concentrate on recent work which bears on 
the census of metals at high redshift (point (b) above).
The current situation is summarised in Figure 1, where the abundances 
measured in different components of the high redshift universe are 
plotted as a function of the column density of neutral gas.
In a rough, statistical sense, higher values $N$(H~I) 
sample regions closer to sites of active star 
formation. The immediate impression from Figure 1 is that our knowledge 
of abundances at $z \simeq 3$ is still very incomplete and undoubtedly
one of the tasks for the next decade will be to fill in the many gaps 
evident in the Figure, as I now discuss.
%
%
\begin{figure}
\vspace*{-2.7cm}
\hspace*{-4.2cm}
\psfig{figure=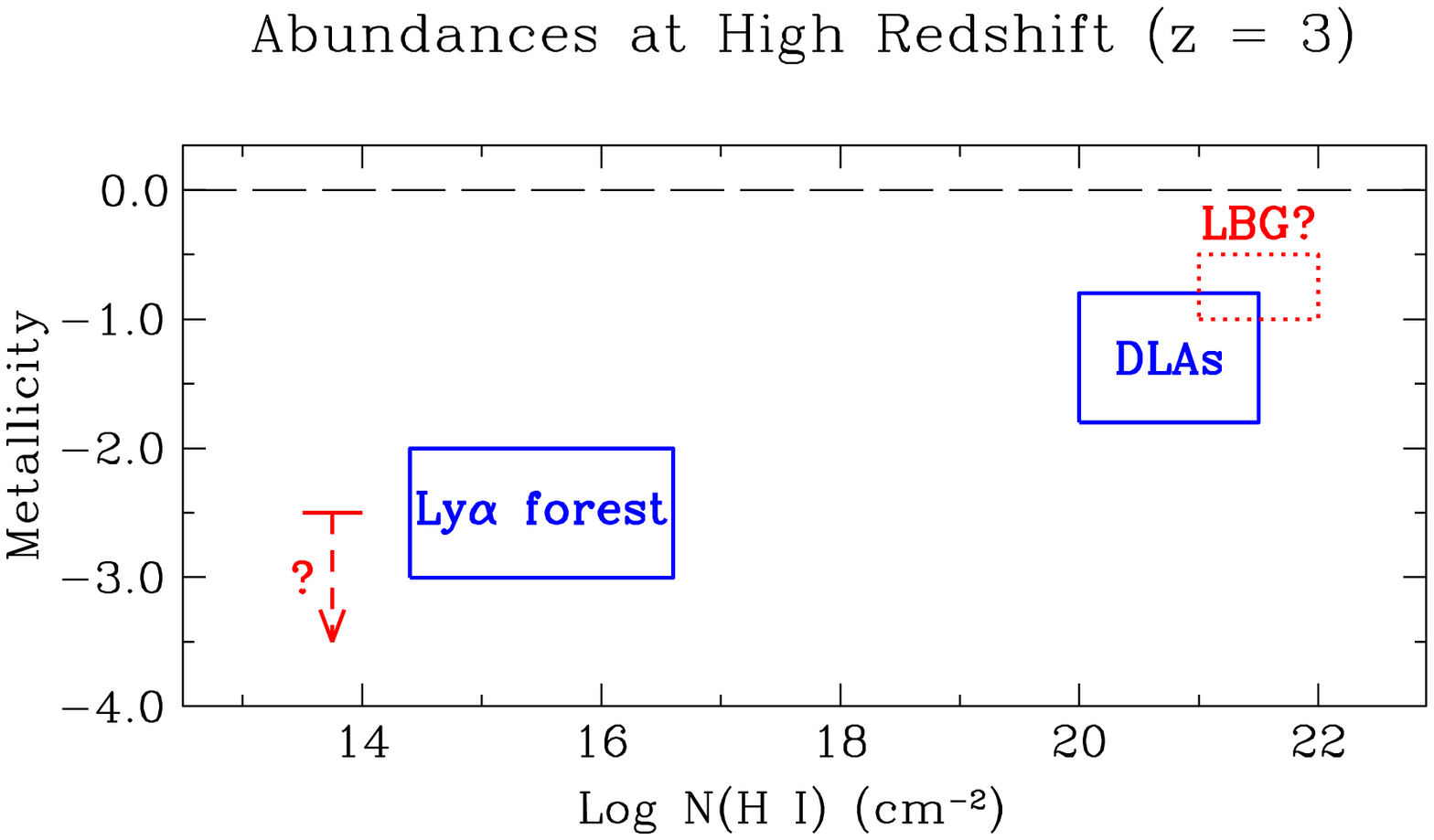,width=14cm,height=20cm,angle=270}
\vspace*{-4.3cm}
\caption{Summary of our current knowledge of abundances at high redshift.
Metallicity is on a log scale relative to solar, and $N$(H~I) is the column
density of neutral hydrogen measured in the \lya\ forest, Damped \lya\ systems
and Lyman break galaxies.}
\end{figure}

\section{The Lyman alpha Forest}
All QSO spectra show a multitude of \lya\ absorption lines 
at redshifts between $z_{\rm em}$ and 0\,.
The distribution of
$N$(H~I) is a power law with slope 
$\beta \simeq 1.7$ (Petitjean et al. 1993); lines corresponding to 
$N$(H~I)$ \leq 10^{17}$~cm$^{-2}$ are normally referred to as the \lya\ 
forest. In the first quantitative paper on the subject, Sargent et al. 
(1980) identified the forest with a population of  
primordial hydrogen `clouds' in pressure equilibrium with a hotter,  
expanding intergalactic medium. 
This picture has evolved more recently  
to one in which the \lya\ forest {\em is} the intergalactic medium, 
based on the results of numerical simulations which show that 
\lya\ absorption 
arises naturally as a consequence of the growth of structure in the 
universe in the 
presence of an ionising background
(e.g. Weinberg, Katz, \& Hernquist 1998).
It is likely that the \lya\ forest accounts for a significant fraction of 
all baryons at $z \simeq 3$, perhaps as much as $\approx 50 - 100$\%
(Rauch et al. 1997; Weinberg et al. 1997).

One of the first discoveries made possible by the High Resolution 
Spectrograph (HIRES, Vogt 1992) on the Keck I telescope was that
the gas producing the \lya\ forest is {\it not} pristine;
approximately 50\% of \lya\ lines with 
$N$(H~I)$ \geq 3 \times 10^{14}$~cm$^{-2}$
and nearly all lines with 
$N$(H~I)$ \geq 1 \times 10^{15}$~cm$^{-2}$
have associated C~IV~$\lambda\lambda1548, 1550$ absorption
(Tytler et al. 1995; Cowie et al. 1995; Songaila \& Cowie 1996).
These lines are simply too weak to have been detected with 
4-m telescopes. 

The abundance of C implied by these detections is however somewhat 
uncertain, primarily because neither H~I nor C~IV are the main ion
stages of hydrogen and carbon at the temperatures and densities thought 
to apply in \lya\ forest. The ionisation corrections for 
hydrogen in particular are between $10^2$ and $10^4$, making any 
inferences on the metallicity highly model dependent. 
A working value  
[C/H] $\simeq -2.5$ is commonly adopted (where as usual
[X/H] = log
[$N$(X)/$N$(H)] $-$ log [$N$(X)/$N$(H)]$_{\odot}$),
but there is probably an order of magnitude dispersion
about this mean within the forest  
(Hellsten et al. 1997; Rauch, Haehnelt, \&  Steinmetz 1997;
Boksenberg, Sargent, \& Rauch 1998).

Irrespectively of the precise value of metallicity, these observations 
show that the products of stellar nucleosynthesis
are widely distributed even at these early epochs
($z \simeq 3$ corresponds to a look-back time of $\approx 90$\% of the age 
of the universe in an Einstein-de Sitter cosmology).
It is unclear, however, whether the source of metals is relatively 
local to the \lya\ lines observed, or whether the
heavy elements have a more widespread, 
and presumably earlier, origin possibly 
associated with a Population III episode of star 
formation. The answer may be found in the metallicity
of the lowest column density \lya\ clouds, which in hydrodynamical 
simulations are located preferentially in regions 
of lower than average matter density. 

In the last year two teams of HIRES observers
have investigated this topic,
with conflicting conclusions. Lu et al. (1998)
co-added the C~IV regions corresponding to almost 300
\lya\ lines 
but still found no composite 
absorption despite the high signal-to-noise ratio achieved
in the stacked data 
(S/N$ \simeq 1860$). 
They concluded that there is a sudden downturn in the 
abundance of C at low column 
densities ([C/H]$\leq -3.5$ when
$N$(H~I)$ \leq 1 \times 10^{14}$~cm$^{-2}$), as predicted by most models
put forward to explain the enrichment of the intergalactic medium.
Cowie \& Songaila (1998), on the other hand, showed that on a 
pixel-by-pixel basis the distributions of C~IV and \lya\ 
optical depths have medians which 
are roughly in a constant ratio
all the way down to 
values as low $\tau$(\lya)$\simeq 0.5 - 1$\,. 
The implication is that whatever
mechanism is at play for the transportation of
heavy elements from the stars that produced them to the lowest density 
regions of the IGM, its efficiency is far higher than anticipated.

Most recently, Ellison et al. (1999) have re-examined this problem 
capitalising on the exceptional brightness ($R = 15.2$) of the newly 
discovered QSO APM~08279+5255 which allowed HIRES observations at 
S/N $\simeq 80$. With the help of 
simulations Ellison et al. showed that 
inevitable differences  
in redshift and velocity dispersion between \lya\ and C~IV absorption
complicate the interpretation of both previous studies. 
It appears that
the question of whether there is a uniform degree of metal 
enrichment in the \lya\ forest 
down to the lowest values of $N$(H~I)
has yet to be settled and will probably 
require even better data than obtained up to now.

\section{Damped Lyman alpha Systems}
Among the different components of the high-$z$ universe this  
is the most favourable to the investigation of chemical abundances.
Over the last few years
Keck observations of damped systems have produced measurements
of exquisite precision (10--20\% in the best cases)
of a wide variety of elements
from carbon to zinc (e.g. Lu et al. 1996; Prochaska \& Wolfe 1999), 
realizing one of the cornerstones of the scientific cases made 
in the 1980s for the construction of 8-m class telescopes.
DLAs are QSO absorption systems at the upper end of the 
neutral hydrogen column density distribution. 
Consequently, 
$N$(H~I) can be  accurately 
determined from the 
profile of the damping wings of the \lya\ absorption line. 
Furthermore,
since the gas is mostly neutral, 
corrections for unobserved ions are
not a major concern---most elements are concentrated in the lowest 
stage whose ionisation potential exceeds that of H~I, as is the case 
in local H~I clouds. 
The main limitation for abundance studies is in accounting for the 
generally unknown and variable fraction of an element which 
has been removed from the gas phase to form interstellar dust
(and therefore does not produce line absorption).
By targeting Zn, which is not significantly depleted onto dust, Pettini et al.
(1997) established that typical abundances at $z \simeq 2 - 3$
are $\approx 1/13$ of solar and that there is almost a two order of magnitude
range in the level of metallicity attained by different damped systems at
essentially the same cosmic epoch.

%
%
\begin{figure}
\vspace*{-1.5cm}
\hspace*{-1.3cm}
\psfig{figure=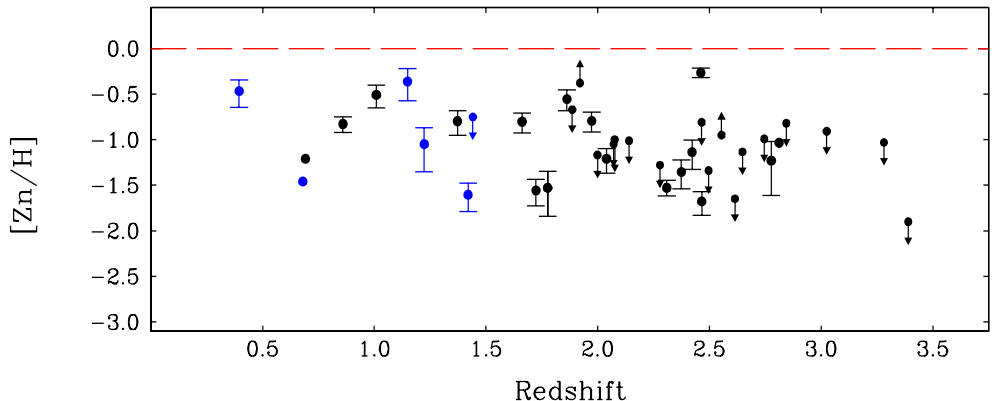,width=14cm,height=20cm,angle=0}
\vspace*{-12.5cm}
\caption{Plot of the abundance of Zn against redshift for 40 DLAs from 
the compilation by Pettini et al. (1999a) which includes 10 systems at 
intermediate redshifts ($z_{\rm abs} = 0.4 - 1.5$). 
No evolution in the column density weighted metallicity
of DLAs can be discerned in the present data set.}
\end{figure}

It is generally assumed, at least as a working hypothesis,
that in DLAs we see the progenitors of present day disk galaxies,
observed at a time when most of their mass was still in gaseous form
(e.g. Wolfe 1990). This connection rests mainly on two pieces of 
circumstantial evidence. First, spiral galaxies 
apparently contribute most of the H~I cross-section in the nearby 
universe (Rao \& Briggs 1993; Natarajan \& Pettini 1997; Zwaan 1998),
although it is important to bear in mind that
this conclusion is drawn from the results of surveys in 21~cm 
emission, rather than damped \lya\ absorption. 
Second, 

\begin{equation}
\Omega_{\rm DLA}{\rm (}z = 3{\rm )} \simeq 0.0025
\approx \Omega_{\rm disk~stars} {\rm (}z = 0{\rm )}
\approx 5 \times \Omega_{\rm H~I}{\rm (}z = 0{\rm )}
\end{equation}

\noindent (e.g. Fukugita, Hogan, \& Peebles 1998). 
It has been tempting on the basis of these rough equalities
to suppose that over the period from $z = 3$ to the present
the gas we see in the damped \lya\ systems turns into
today's stars, leaving a gaseous fraction of $10-20$\% in galaxies
like the Milky Way. 
The caveats here are that all three values of $\Omega$ above are only 
approximately known and, most importantly, they are only a small fraction 
of the baryons available;
$\Omega_{\rm DLA}$ $\approx$  $\Omega_{\rm disk~stars}$ $< 1/20 \Omega_B$\,.
More recently it has been argued that the profiles of the metal 
absorption lines in DLAs are indicative of large rotating disks
(Prochaska \& Wolfe 1997), but even this line of evidence is  
open to alternative interpretations 
(Haehnelt, Steinmetz, \& Rauch 1998; Ledoux et al. 1998;
Rauch, Sargent, \& Barlow 1999).

The working model of DLAs as the progenitors of today's spirals makes 
three straightforward predictions as we follow the evolution of these QSO 
absorbers from high redshift to the present time:
(1) Direct imaging should reveal bright galaxies close to 
the QSO sight-lines; (2) $\Omega_{\rm DLA}$ should decrease from 
$\Omega_{\rm DLA}$($z = 3$) to
$\Omega_{\rm H~I}$($z = 0$); and (3) the metallicity $Z_{\rm DLA}$
should increase from $1/13 Z_{\odot}$ (the value measured at $z = 3$)
to $\sim Z_{\odot}$ (Edmunds \& Phillipps 1997). Points (2) and (3) in 
particular have 
been central to models of global chemical evolution 
(e.g. Pei \& Fall 1995).
{\it None} of these predictions has been borne out by observations of DLAs
at $z \simlt 1$, made possible in the last couple of years by the 
slowly increasing archive of {\it HST} ultraviolet spectra.
Addressing the three points above in order:
(1) {\it HST} imaging has shown galaxies of different 
morphological types and luminosities
to be associated with DLAs
(Le Brun et al. 1997; Lanzetta et al. 1997; Rao \& Turnshek 1998);
(2) $\Omega_{\rm DLA}$($z < 1$) $\approx$ $\Omega_{\rm DLA}$($z > 2$)
so that there is no evidence for any redshift evolution 
in $\Omega_{\rm DLA}$ (Turnshek 1998); and 
(3) Similarly,  
$Z_{\rm DLA}$($z < 1$) $\approx$ $Z_{\rm DLA}$($z > 2$),
as can be seen from Figure 2.

The current body of data thus points to the conclusion
that known damped \lya\ systems trace a diverse population 
of which the galaxies responsible for the bulk of the star formation 
are {\it not} the major component, at least at intermediate and low 
redshifts. With hindsight, there are probably good reasons for this.
As emphasized most recently by Mo, Mao, \& White 
(1999---see also Jimenez, Bowen, \& Matteucci 1999),
given a range of surface densities and a Schmidt law of star formation,
the most compact galaxies will be the sites of rapid star formation, while
the more diffuse, and more slowly evolving, systems will dominate the 
absorption cross-section. Even so, the point stressed originally by Pei 
\& Fall (1995) still applies: a column 
density-weighted estimate of $Z_{\rm DLA}$ remains a valid
measure of the metallicity of the neutral component of the universe,  
irrespectively of the exact morphological mix,
{\it provided no particular class of galaxy is actually excluded from
the census}. It is certainly possible
that current samples of DLAs miss the most metal rich
galaxies, simply because any QSOs which happen to lie behind them (as 
viewed from Earth) would suffer sufficient dust extinction 
to be excluded 
from magnitude limited samples. This bias is likely to be most 
significant at $z < 1.5$, given the relatively small aperture
of {\it HST} (the only instrument which can record a damped \lya\ 
absorption line at wavelengths below 3050~\AA). It remains to be seen how 
severe an effect dust bias really is at $z = 2 - 3$\,. 

\section{Lyman Break Galaxies}
Large telescopes have now brought within our reach 
{\it direct} spectroscopic studies of high redshift galaxies. 
The success of the Lyman break technique 
has exceeded even the most optimistic 
expectations (e.g. Steidel \& Hamilton 1992), with the current tally 
standing at more than 600 galaxies with measured redshifts $z > 2.5$
whereas none were known only three years ago
(Steidel et al. 1998a).
The technique is essentially a very efficient photometric selection.
By targeting the inherent spectral discontinuity due to the ionization 
edge of hydrogen at a rest wavelength $\lambda_0 = 912$~\AA, 
it is possible to design volume limited surveys 
within well defined redshift intervals 
determined by one's choice of filters. 

%
%
\begin{figure}
\vspace*{-0.5cm}
\hspace*{-0.8cm}
\psfig{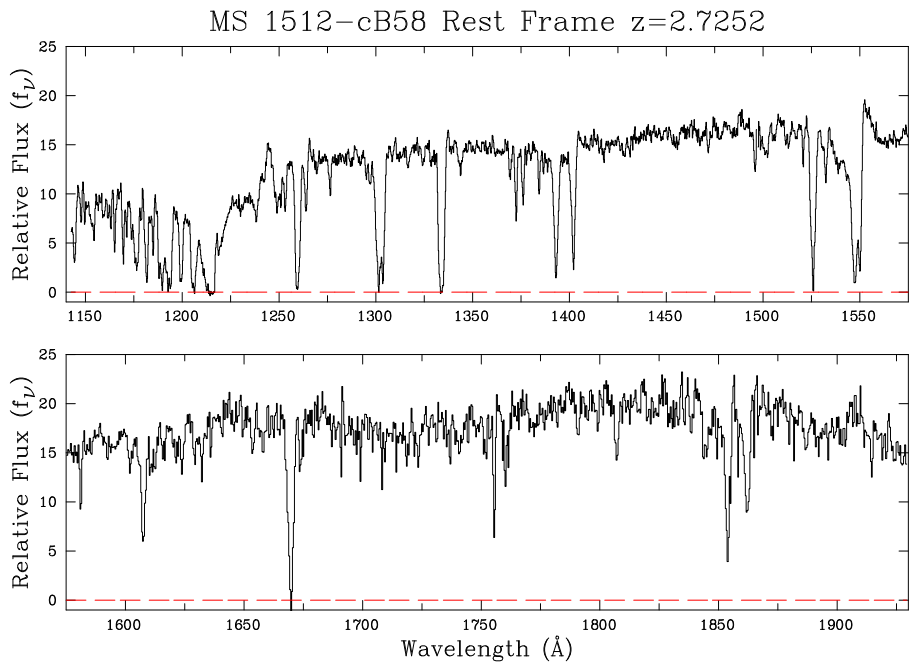}
\caption{Spectrum of the $V = 20.64$, $z = 2.7252$
galaxy MS 1512-cB58 obtained with LRIS on the
Keck~I telescope in May and August 1996. In the blue spectrum (top panel) 
we reached S/N $\simeq 50$ per pixel at a resolution of 3.0
\AA\ FWHM with a total exposure time
of 11\,400~s. The strongest absorption lines are mostly of interstellar 
origin. (Pettini et al. 1999b, in preparation).}
\end{figure}

By selecting objects on the basis of their rest frame ultraviolet 
colours, the Lyman break method naturally picks out galaxies with 
on-going star formation activity. It is therefore no surprise to find 
that their spectra resemble closely
those of local starburst galaxies and
are characterised by a blue continuum, strong stellar and interstellar 
absorption lines 
and generally weak \lya\ emission.
Lyman break galaxies are clearly as close as we 
get to the production sites of heavy elements  
and it is here above all else that we would like to 
focus our abundance studies. 
It is therefore somewhat ironical that, notwithstanding
the exceptionally high quality of some of the data
(Figure 3 is the best example), their rich UV spectra 
are singularly ill-suited to accurate abundance determinations.

The large equivalent widths of the saturated interstellar lines 
($W_0 \simeq 2 - 4$~\AA\ in Figure 3)
are an indication that the interstellar media of these galaxies are 
stirred to large velocity dispersions---presumably by the high
supernova rate which must accompany the star formation activity we 
see, and do not directly measure the metal content of the gas. 
It is true that in local starburst galaxies there appears to be  
a rough correlation 
between these equivalent widths and the underlying 
abundances (Heckman et al. 1998). 
However it is unclear whether this trend, which ultimately has its 
origin in the metallicity -- luminosity relation, 
also applies at high redshift and, in 
any case, it only gives a rough statistical measure of metallicity.
Stellar photospheric lines do not help either, as they are more sensitive 
to temperature and density than abundance. 
The terminal velocity of high ionisation lines produced in the 
expanding atmospheres of O and B supergiants may be more promising, since 
metallicity does play an important part in the coupling of radiative and 
mechanical energies, but a proper calibration has yet to be established
(see Claus Leitherer's contribution to these proceedings for a detailed 
discussion of all these issues).

%
%
\begin{figure}
\vspace*{-3.5cm}
\hspace*{-2cm}
\psfig{figure=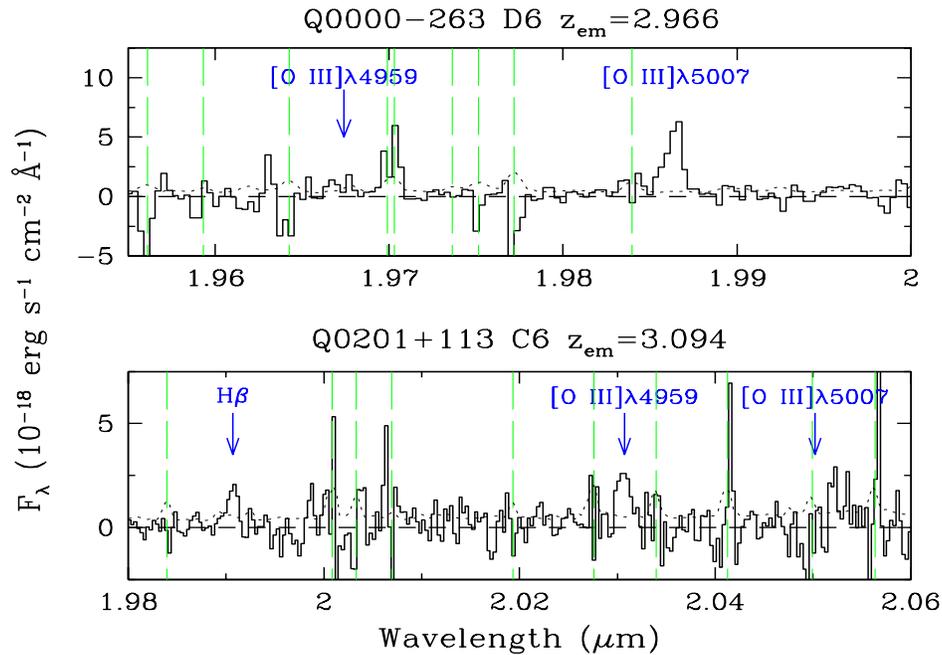,width=17cm,height=20cm,angle=0}
\vspace{-8.0cm}
\caption{Examples of {\it UKIRT} near-IR spectra of Lyman break galaxies, 
reproduced from Pettini et al. (1998).
The positions of the nebular emission lines covered are indicated.
The short-dash line shows the $1 \sigma$ error applicable to each
spectrum. The vertical long-dash lines mark the locations of the strongest
sky OH emission features; although they have been subtracted out,
large residuals can remain in cases where the sky lines are saturated.
The exposure times were 21\,600 and 18\,000 s for Q0000$-$263~D6 and
Q0201$+$113~C6 respectively.}
\end{figure}

In my view the best prospects for determining the degree of metal 
enrichment of Lyman break galaxies 
are offered by the familiar $R_{23}$ method 
which relates the strength of [O~II] 
and [O~III] emission to H$\beta$ 
(Pagel et al. 1979; Kobulnicky, Kennicutt, \& Pizagno 1999).
At $z \simeq 3$ these lines are redshifted into the near infrared, but 
with a large sample of objects it is possible to select redshifts which 
place them in gaps between the much stronger OH sky lines.
Pilot observations with CGS4 on UKIRT have demonstrated the feasibility 
of this approach (see Figure 4); we now await the availability of 
near-IR high resolution spectrographs on large telescopes, such as ISAAC 
in the VLT, to exploit the method in full.
Ultimately, we shall need {\it NGST} for a full 
coverage of the nebular spectra from the H~II regions of these distant 
galaxies (Kennicutt 1998).

\section{Epilogue: Where are the Metals at High Redshift?}
I suspect that one's reaction to the work reviewed above will depend 
very much on one's background. 
Cosmologists may be excited by the mere fact that we can 
begin to address observationally the whole question of chemical 
abundances in the distant universe, when 10 years ago
this was an area beyond our reach.
On the other hand, astronomers who for many years
have been accustomed to measuring abundances in stars and nearby galaxies 
with a great deal of precision will rightly view
the results I have presented as rudimentary
and preliminary. 
Given the gaps in our knowledge so evident from Figure 1, 
it may seem somewhat foolish to even attempt
a census of metals at high redshift, but I will go ahead with it anyway!

%
%
\begin{figure}
\vspace*{-2.5cm}
\hspace*{-4.2cm}
\psfig{figure=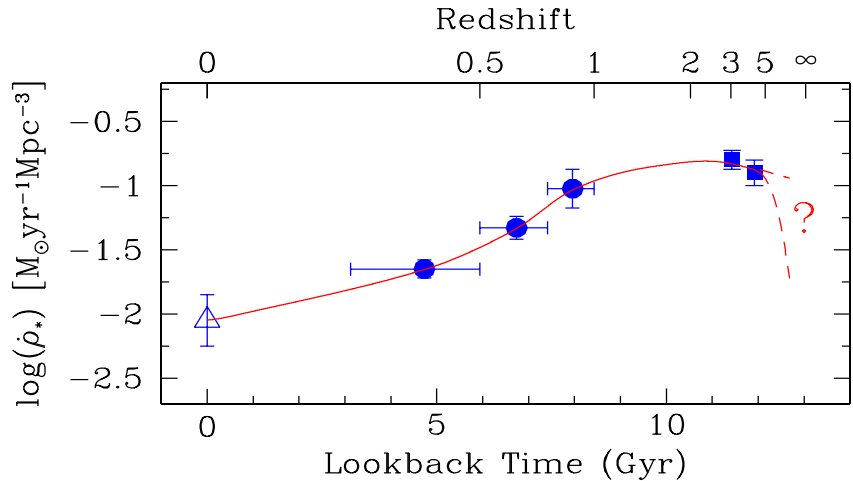,width=14cm,height=20cm,angle=270}
\vspace*{-4.5cm}
\caption{The comoving star formation rate density 
$\dot{\rho}_{\ast}$ vs.
lookback time compiled from wide angle, ground based
surveys (Steidel et al. 1999a and references therein).
The data shown are for a $H_0 = 50$~km~s$^{-1}$~Mpc$^{-1}$,
$q_0 = 0.5$ cosmology.
}
\end{figure}

Figure 5 shows the most recent version of a 
plot first constructed by Madau 
et al. (1996) which attempts to trace the `cosmic star formation 
history' by following the redshift evolution of the 
comoving luminosity density of star forming 
galaxies.
Over the past two years this plot has changed considerably.
The star formation rates derived from UV continuum luminosities
(filled symbols in Figure 5) have increased by factors 
of $\sim 3$ and $\sim 5$ at 
$z \leq 1$ and $z \geq 3$ respectively to account for attenuation of the 
emergent starlight by dust (these corrections have been deduced by 
comparing SFRs derived from the UV continuum and from
Balmer lines---Tresse \& Maddox 1998; Glazebrook et al. 1999, 
Pettini et al. 1998). 
Furthermore, there is no longer evidence for
an epoch of enhanced star formation at $z \simeq 1 - 2$, but rather
the best estimates available at present favour 
a roughly constant star formation activity 
between $z \simeq 1$ and $\sim 4$, and probably push
the end of the `dark ages', when the universe became reionised,
to $z \simgt 5$.
The decline in $\dot{\rho}_{\ast}$ from $z = 3$ to 4 
reported by Madau et al. (1996) now 
appears to have been due
to cosmic variance in the small volumes probed by the 
{\it Hubble Deep Field} (Steidel et al. 1999a).

These changes have highlighted the many uncertainties which still apply 
to the plot in Figure 5, such as unquantified selection effects in the 
way high-$z$ galaxies are picked out in the first place; 
incomplete knowledge of the luminosity 
function at different redshifts, particularly at the faint end; 
and total ignorance of the 
IMF and dust extinction law which apply to these early episodes of star 
formation.
It is therefore likely that Figure 5 will undergo further revisions
as future surveys improve 
our knowledge of distant galaxies (e.g. Cowie, Songaila, \& Barger 1999).   
However, if we assume for a moment that we have the story about right, 
some interesting consequences follow.

The first question one may ask is: {\it ``What is the total mass in stars
obtained by integrating under the curve in Figure 5?''}
The data points in Figure 5 were derived assuming a Salpeter IMF
(with slope $-1.35$) between $M = 100~M_{\odot}$ and $0.1~M_{\odot}$\,.
However, Leitherer (1998) has argued that the  
scaling normally used to convert the observed UV luminosity
into a SFR overestimates the SFR by a factor of 2.5
if, more realistically, the IMF flattens below 
$1~M_{\odot}$\,. 
With Leitherer's correction,\footnote{We 
have not applied this correction directly to 
Figure 5 in order not to confuse the comparison with earlier versions of 
this plot.} 

\begin{equation}
	\int_{0}^{13~Gyr}\dot{\rho}_{\ast}\hspace{-0.8mm}'~dt \simeq
	3.3 \times 10^8~M_{\odot}~{\rm Mpc}^{-3} =
	0.0043~\rho_{\rm crit} \approx \Omega_{\rm stars}~\rho_{\rm crit}
	\label{}
\end{equation}

\noindent where 
$\rho_{\rm crit} = 7.7 \times 10^{10}~h_{50}^2~M_{\odot}~{\rm Mpc}^{-3}$
and $\Omega_{\rm stars}$ is the fraction of the closure density
contributed by stars at $z = 0$ (Fukugita et al. 1998).
Thus, within the rough accuracy with which this accounting can 
be done, the star formation history depicted in Figure 5 is apparently 
sufficient to produce the entire stellar 
content, in disks and spheroids, of the present day universe.
Note also that $\approx 1/4$ of today's stars were
made before $z = 2.5$ (the uncertain 
extrapolation of $\dot{\rho}_{\ast}$ beyond $z = 4$ 
makes little difference).

We can also ask: {\it ``What is the total mass of metals produced by $z = 
2.5$?''} (this being the redshift at which the most extensive set of 
abundance measurements at high $z$ is available).
Using the conversion factor 
$\dot{\rho}_{\rm metals} = 1/42~\dot{\rho}_{\ast}$
to relate the comoving density of ejected metals to
the star formation rate density (Madau et al. 1996)
we find 

\begin{equation}
	\int_{11~Gyr}^{13~Gyr}\dot{\rho}_{\rm metals}~dt \simeq
	4.5 \times 10^6~M_{\odot}~{\rm Mpc}^{-3} \simeq
	0.04 \times {\rm (}\Omega_B \times 0.0189 {\rm )}
	\label{}
\end{equation}

\noindent where for the density of baryons we adopt 
$\Omega_B = 0.076~\Omega$ (\S 1 above)
and 0.0189 is the mass fraction of elements heavier than helium
for solar metallicity
(Anders \& Grevesse 1989).
In other words, the amount of metals produced by the star formation we 
{\it see} at high redshift (albeit corrected for dust extinction)
is sufficient to enrich the whole baryonic content of the universe 
at $z = 2.5$ to $\approx 1/25$ of solar metallicity.
Note that this conclusion does not depend sensitively on the IMF.
%
%
\begin{figure}
\vspace*{-6.8cm}
\hspace*{-3.3cm}
\psfig{figure=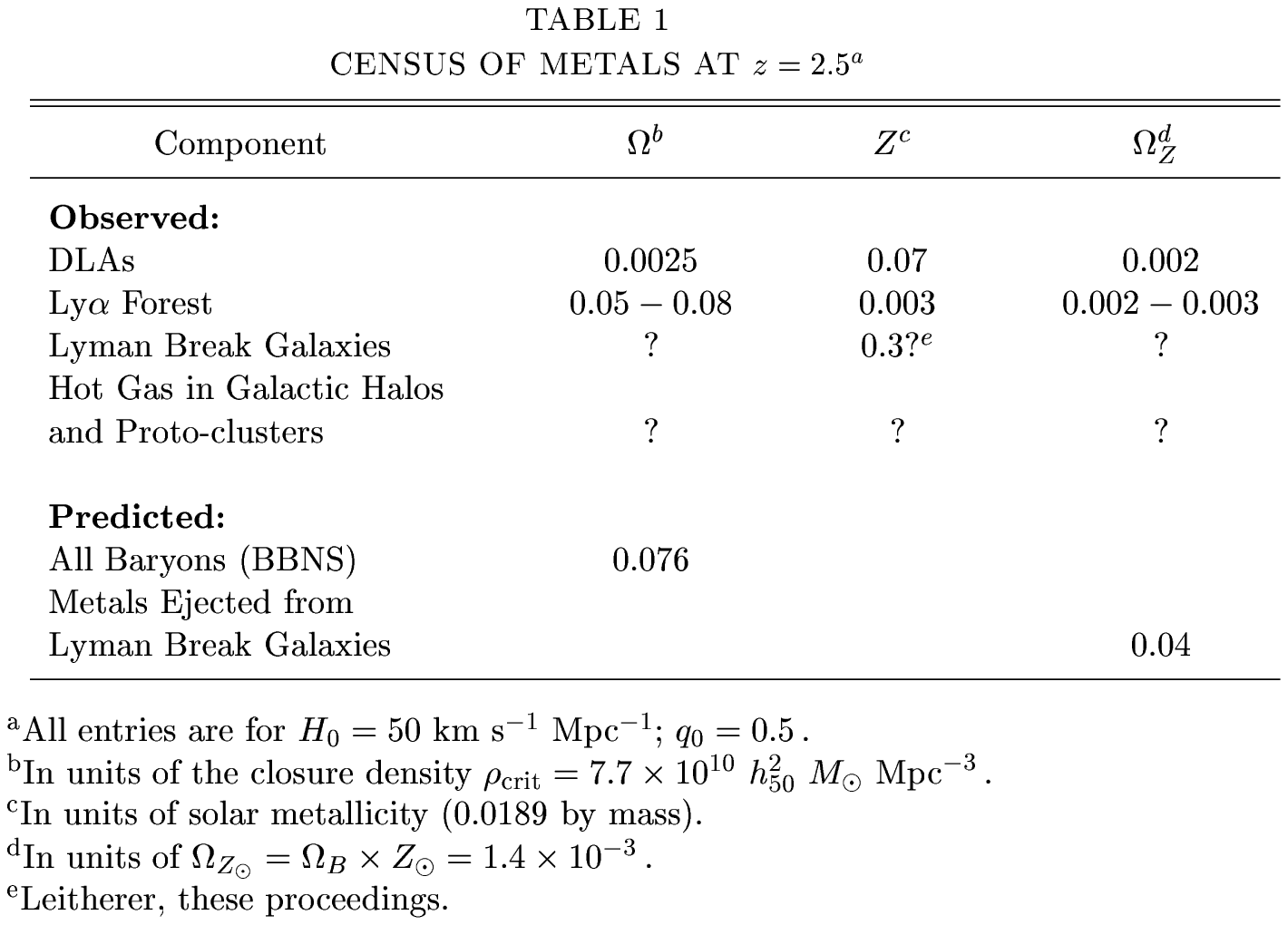,width=17.5cm,angle=0}
\vspace{-9cm}
\end{figure}

As can be seen from Table 1, this leaves us with a serious 
`missing metals' problem which has already been noted by Pagel (1998).
The metallicity of damped \lya\ systems is in the right ball-park,
but $\Omega_{\rm DLA}$ is only a small fraction of $\Omega_B$. Conversely, 
while the \lya\ forest may account for a large fraction of the baryons, 
its metal content is one order of magnitude too low.
Therefore, when we add up all the metals which have been measured
with some degree of confidence up to now, we find that they account
for no more than $\approx 10$\% of what we expect to have been produced 
and released by $z = 2.5$ (last column of Table 1). 
Lyman break galaxies are unlikely to make up 
the shortfall, even though their metallicity may already be high at 
these redshifts. However, large-scale outflows of metal-enriched material
are commonly seen in these galaxies (e.g. Pettini et al. 1998), and it is 
therefore possible that the missing metals are to be found far from 
their production sites, in hot gas in galactic 
halos and proto-clusters, as argued for example by Renzini (1998).
Such gas has already been seen in both 
absorption and emission (Kirkman \& Tytler 1997, 1999; Steidel et al. 
1998b; Steidel et al. 1999b, in preparation), although we have no idea yet 
of its metallicity and baryon content. 
Filling in this major gap in the metal budget at high redshift is 
undoubtedly one of the priorities for abundance studies as we enter the 
VLT era.\\

I am very grateful to the organisers of this most fruitful meeting for 
inviting me, to
my collaborators in the various projects described in this article,
and to Bernard Pagel and Mike Edmunds for illuminating discussions.

%
%

\end{document}